\documentclass{revtex4}

\def\bea{\begin{eqnarray}}
\def\eea{\end{eqnarray}}
\def\bec{\begin{center}}
\def\ec{\end{center}}

\def\beq{\begin{equation}}

\def\f{\frac}

\def\f#1#2{\frac{#1}{#2}}

\def\pr{\prime}
\def\l{\left}
\def\r{\right}
\def\a{\alpha}
\def\t{\tilde}

\def\Lm{\Lambda}

\def\bea{\begin{eqnarray}}
\def\eea{\end{eqnarray}}
\def\bec{\begin{center}}
\def\ec{\end{center}}

\def\f#1#2{\frac{#1}{#2}}

\def\l{\left}
\def\r{\right}
\def\a{\alpha}
\def\Lm{\Lambda}

\def\t{\tilde}

\def\f{\frac}

\def\beq{\begin{equation}}
\def\eeq{\end{equation}}

\def\beq{\begin{equation}}
\def\eeq{\end{equation}}

\begin{document}
\draft
\tighten
\preprint{KAIST-TH 03/??}
\title{\large \bf One-loop Gauge Couplings
in Orbifold Field Theories}
\author{Kiwoon Choi\footnote{kchoi@hep.kaist.ac.kr}
and Ian-Woo Kim\footnote{iwkim@hep.kaist.ac.kr}}
\affiliation{Department of Physics, Korea Advanced Institute of 
Science and Technology\\ Daejeon
305-701, Korea}
\date{\today}
\begin{abstract}
We discuss the gauge coupling renormalization
in orbifold field theories in which 
the 4-dimensional graviton and/or matter fields
are quasi-localized in extra dimension
to generate hierarchically different mass scales
and/or Yukawa couplings.
In such theories, there can be large calculable Kaluza-Klein
threshold corrections to low energy gauge couplings,
enhanced by the logarithms of small
warp factor and/or of small Yukawa couplings.
We present the results on those Kaluza-Klein threshold corrections
in generic 5-dimensional theory on $S^1/Z_2\times Z_2$
containing arbitrary 5-dimensional gauge, spinor and scalar fields.
\end{abstract}
\maketitle

\section{introduction}

It has been noticed that theories with extra dimension 
can provide an elegant mechanism to generate various hierarchical 
structures in 4-dimensional (4D) physics, e.g.
the weak to Planck scale hierarchy $M_W/M_{Pl}\approx 10^{-16}$
\cite{add,rs}
and the hierarchically different Yukawa couplings ranging from
$y_e\approx 10^{-6}$ to $y_t\approx 1$
\cite{yukawa}.
A particularly interesting mechanism to generate the scale
hierarchy $M_W/M_{Pl}\approx 10^{-16}$ is the quasi-localization
of 4D graviton in extra-dimension \cite{rs}.
In 5D theory on $S^1/Z_2$ which is parameterized by 
$y=[0,\pi]$, if the 5D cosmological constant is
negative and the brane cosmological constants are appropriately tuned, 
the resulting geometry is a slice of $\mbox{AdS}_5$, yielding
a 4D graviton quasi-localized at $y=0$ with wavefunction $e^{-kRy}$
where $k$ is the AdS curvature and $R$ is the radius of $S^1$.
If the Higgs boson for electroweak symmetry
breaking is assumed to be confined at the orbifold fixed point
$y=\pi$, 
its wavefunction-overlap with quasi-localized graviton is suppressed 
by the warp factor $e^{-kR\pi}$, leading to an exponentially small
scale ratio $M_W/M_{Pl}\approx e^{-\pi kR}$. 
One can obtain also the hierarchically different 
Yukawa couplings by quasi-localizing fermions in extra dimension
\cite{yukawa}.
Again for the Higgs boson confined at $y=\pi$, fermions
quasi-localized at $y=\pi$ have Yukawa couplings of
order unity. On the other hand, fermions quasi-localized at 
the other fixed point $y=0$
have an exponentionally small wavefunction-overlap with
the Higgs boson, and thus small Yukawa couplings.

Grand unification of the strong and electroweak forces
is a highly persuasive idea for physics at high
energy scales.
However conventional 4D grand unified theories (GUTs) have suffered
from well-known problems such as the doublet-triplet splitting
problem and the issue of too rapid proton decay.
GUTs in higher dimensional spacetime can avoid these problems 
through the mechanism of symmetry breaking by 
boundary conditions \cite{kawamura}.
It is straightforward to implement the idea of 
quasi-localization in orbifold GUTs to generate
the scale and/or Yukawa hierarchies
\cite{hebecker}.

In any GUT, heavy particle threshold effects at GUT-symmetry breaking scale 
should be taken into account for a precision analysis of
low energy gauge couplings.
In conventional 4D GUT,  those heavy particle threshold corrections 
are {\it not} so important since they are subleading compared to
the leading large log effects from light fields.
However orbifold GUT contains
(infinitely) many Kaluza-Klein (KK) modes,
thus can have sizable GUT-scale threshold corrections 
\cite{choi}.
There are two different type of KK threshold corrections in orbifold GUT:
one which is power-law divergent \cite{ddg} and the
other which is either log-divergent or finite
\cite{contino,hall}.
The power-law divergent parts are sensitive to the unknown UV completion,
and thus {\it not} calculable within orbifold
field theory
\cite{contino,choi1,powerlaw}.
Still the GUT-symmetry in bulk spacetime
guarantees that the power-law divergent parts are universal for
all gauge couplings, so do {\it not} affect the gauge coupling 
differences which are of phenomenological interests.
The log-divergent or finite parts are calculable within orbifold GUT,
and not universal in general since
the GUT-symmetry is broken by boundary conditions.
As we will see, these calculable threshold corrections
are enhanced by
the {\it large} logarithm of an exponentially small warp factor and/or of 
an exponentially small Yukawa coupling
when the scale and/or Yukawa hierarchies are generated by
quasi-localization \cite{pomarol,choi1,goldberger,choi2,choi3}, 
and thus should be taken into account
in the analysis of low energy gauge couplings.

The aim of this talk is to discuss and summarize the results
on  KK threshold corrections in generic 5D orbifold field theories
in which the scale and/or Yukawa hierarchies
are generated by quasi-localization \cite{choi1,choi2,choi3}.
In Section II, we discuss briefly how the quasi-localization
of graviton and matter fermions lead to the scale and Yukawa
hierarchies.
In Section III, we present the results on
KK threshold corrections in generic
5D gauge theory on a slice of $\mbox{AdS}_5$.
In Section IV, we discuss an alternative way to compute
KK threshold corrections in supersymmetric theories,
which is based on 4D effective supergravity (SUGRA).
In Section V, we briefly summarize the results for 5D theories
on flat extra dimension in which matter fermions are
quasi-localized to generate hierarchical Yukawa couplings.

\section{Scale and Yukawa hierarchies from quasi-localization}

Since the mass scales in 4D physics are measured by 4D graviton,
a dynamical quasi-localization of 4D graviton in extra dimension
can generate hierarchically different mass scales in 4D physics.
To see this, let us consider the Randall-Sundrum model \cite{rs}
on $S^1/Z_2$ with 5D action:
\beq
S=M_5^3\int\, d^5x\,
\sqrt{-G}\,\l[-\f{1}{2}{\cal R}+
6k^2-\frac{1}{\sqrt{G_{55}}}\l\{
\delta(y)6k-\delta(y-\pi)\l(6k-G^{\mu\nu}D_\mu H
D_\nu H^*-M_H^2HH^*\r)\r\}\,\r],
\eeq
where $M_5$ and ${\cal R}$ are the 5D Planck scale and 
Ricci scalar for the 5D metric $G_{MN}$, 
$H$ is the Higgs field for electroweak symmetry breaking
which is confined at $y=\pi$,
and the 5D mass parameters $k \,(>0)$ and $M_H$ are assumed 
to be comparable to $M_5$. 
The equations of motion from this action determine
the spacetime geometry to be  a slice of $\mbox{AdS}_5$:
\beq
\label{adsmetric}
ds^2=G_{MN}dx^Mdx^N=e^{-2kRy}g_{\mu\nu}dx^\mu dx^\nu+R^2 dy^2\,,
\eeq
where $g_{\mu\nu}$ corresponds to the massless 4D graviton,
$R$ is the orbifold radius, 
and $y=[0,\pi]$ is the coordinate of $S^1/Z_2$.
This solution shows that the 4D graviton is quasi-localized
at $y=0$ due to the negative 5D cosmological constant.

The Higgs boson mass $M_H$ measured by $G_{MN}$
is generically of the order of $M_5$.
However the observed electroweak scale is measured by
the 4D graviton $g_{\mu\nu}$.
Since the 4D graviton wavefunction at $y=\pi$ 
is exponentialy small, the Higgs boson mass $m_H$ measured
by $g_{\mu\nu}$ is red-shifted by the warp factor as
$$
m_H=e^{-\pi kR}M_H\,.
$$
This can be easily seen by considering the 4D effective action of
$g_{\mu\nu}$ and $H$:
\beq
S_{\rm eff}=\int d^4x \,\sqrt{-g}\,\l[
-\f{M_5^3}{2k}\l(1
-e^{-2\pi kR}\r){\cal R}(g)-g^{\mu\nu}D_\mu HD_\nu H^*
-e^{-2\pi kR}M_H^2HH^*\,\r]
\eeq
yielding the weak to Planck scale ratio measured by the 4D graviton:
\beq
\f{m_H}{M_{Pl}}=\f{e^{-\pi kR}M_H}{\sqrt{M_5^3(1-e^{-2\pi kR})/k}}
\approx e^{-\pi kR}\approx 10^{-16}
\eeq
for $M_5\approx k\approx M_H$ and $kR\approx 12$.
In fact, the above  red-shift applies to all mass scales
at $y=\pi$. For any dimensionful coupling $\lambda_5$ at $y=\pi$
defined in the metric frame of $G_{MN}$, the corresponding
dimensionful coupling
$\lambda_4$ measured by $g_{\mu\nu}$ is red-shifted as
$$
\lambda_4=\l(e^{-\pi kR}\r)^{D_\lambda}\lambda_5\,,
$$
where $D_\lambda$ is the mass-dimension of $\lambda_5$.

Hierarchical Yukawa couplings can be generated similarly
by quasi-localizing matter fermions \cite{yukawa}.
To see this, let us consider a 5D theory on $S^1/Z_2$
containing 5D fermions and also a 4D Higgs field
confined at $y=\pi$:
\beq
S=-\int d^5x \sqrt{-G}\l[ \, 
\,i \bar{\Psi}_I(\gamma^MD_M+
{M}_I\epsilon(y))\Psi_I
+\frac{\delta(y-\pi)}{\sqrt{G_{55}}}
\l(\, D_\mu HD^\mu H^*+
\f{\lambda_{IJ}}{\Lambda} H\psi_I\psi_J\,\r)\,\r]\,.
\eeq
where $\epsilon (y)=y/|y|$,
$\Lambda$ denotes the cutoff scale of 5D
orbifold field theory, and
$\lambda_{IJ}$ are dimensionless brane Yukawa couplings.
Here we assume that the spacetime geometry is flat, so
$$
ds^2=G_{MN}dx^Mdx^N=g_{\mu\nu}dx^\mu dx^\nu+R^2dy^2\,.
$$
The 5D Dirac fermion $\Psi_I$
has the boundary condition
$$
\Psi_I(-y)=z_I\gamma_5\Psi_I(y), \quad
\Psi_I(-y')=z_I \gamma_5\Psi_I(y')\,,
$$
where $y'=y-\pi$, $z_I=\pm 1$,
and
$\psi_{I}=\f{1}{2}(1+\gamma_5)\Psi_{I}$
$(z_I=1)$ or
$\f{1}{2}(1+\gamma_5)\Psi_{I}^c$
$(z_I=-1)$.
For any value of the $Z_2$-odd mass $M_I$,
the 5D fermion $\Psi_I$ has a 4D chiral zero mode
$$
\psi_{0I}=\exp (-z_IM_IRy)\,,
$$
which is quasi-localized at $y=0$ or $\pi$, depending
on the sign of $M_I$.
It is then straightforward to find that the 4D Yukawa couplings
of {\it canonically normalized} fermion zero modes are given by
\beq
\label{canonicalyukawa}
y_{IJ} =\sqrt{Z(z_I{M}_I)Z(z_J{M}_J)}\,\,
\lambda_{IJ}
\eeq
where 
$$
Z(M) =\f{M}{\Lambda}\f{1}{e^{2M\pi R}-1}.
$$
Obviously, the 4D Yukawa couplings
$y_{IJ}$ can have very different values,
depending upon the values of $M_I$,
even when the 5D parameters $\lambda_{IJ}$ 
have similar values.
For instance, for $z_IM_I$ and $z_JM_J\lesssim - 1/R$,
we have
\beq
y_{IJ}\approx \sqrt{\l|\f{M_IM_J}{\Lambda^2}\r|}
\,\,{\lambda}_{IJ}\,,
\eeq
while for $z_IM_I$ and $z_JM_J\gtrsim 1/R$,
\beq
\label{smallyukawa}
y_{IJ}\approx \sqrt{\l|\f{M_IM_J}{\Lambda^2}\r|}\,
e^{-(z_IM_I+z_JM_J)\pi R}\,\,{\lambda}_{IJ}.
\eeq
The physical interpretation of this result is simple.
If $z_{I,J}M_{I,J}\gtrsim 1/R$, 
the corresponding zero modes are quasi-localized at  $y=0$, 
so the Yukawa couplings are exponentially 
suppressed as they originate from $y=\pi$.
On the other hand, for $z_{I,J}M_{I,J}\lesssim
-1/R$, the zero modes are localized at $y=\pi$, so
there is no suppression of Yukawa couplings.

\medskip

\section{one loop gauge couplings in $\mbox{AdS}_5$}

The model we study in this section is a 5D gauge theory defined
on a slice of AdS5 with the spacetime metric (\ref{adsmetric})
\cite{choi2}. 
The lagrangian is given by 
\bea
S=-\int d^4 x d y \sqrt{-G} \left[ \f{1}{4\hat{g}_{5a}^2} F^{aMN} F^a_{MN}
+D_M \phi D^M \phi^* +\hat{m}^2\phi\phi^* 
+i \bar{\Psi} (\gamma^M D_M +M\epsilon(y)) \Psi \right],
\eea
where $D_M$ is the covariant derivative containing the gauge connections 
as well as the spin connection of $\mbox{AdS}_5$.
Here we include the 4D gauge kinetic terms and scalar mass-squares 
confined on the orbifold fixed points as well as the
conventional 5D kinetic and mass terms, so
\bea
\label{mass}
&&\f{1}{\hat{g}_{5a}^2}=
\f{1}{g_{5a}^2}+\f{\delta(y)}{R}\f{1}{g_{0a}^2}
+\f{\delta(y-\pi)}{R}\f{1}{g_{\pi a}^2}\,,
\nonumber \\
&&\hat{m}^2=A^2k^2 +  \f{2k}{R} \left[ \,
B_0 \delta(y) - B_\pi \delta(y-\pi) \,\right],
\quad
M = Ck\,.
\eea
The 5D fields in the model can have arbitrary 
$Z_2\times Z_2^{\prime}$ orbifold boundary condition,
\bea
&&\phi(-y)=z_\phi \phi(y)\,,\quad
\phi(-y')=z^{\prime}_\phi \phi(y')\,,
\nonumber \\ 
&&\Psi(-y)=z_\Psi \gamma_5\Psi(y)\,,\quad
\Psi(-y')=z^{\prime}_\Psi \gamma_5\Psi(y')\,,
\nonumber  \\
&& A^a_\mu(-y)=z_a A^a_\mu(y)\,,\quad
A^a_\mu(-y')=z^{\prime}_a A^a_\mu(y')\,,
\nonumber
\eea
where $z_\Phi,z^{\prime}_\Phi=\pm 1$ for
$\Phi=\{\,\phi,\Psi, A^a_M\,\}$ and $y'=y-\pi$.

The one-loop gauge couplings at low momentum scale $p$
are given by
$$
\f{1}{g_a^2(p)}=\l(\f{1}{g_a^2}\r)_{\rm tree}
+\l(\f{1}{g_a^2}\r)_{\rm loop}\,,
$$
where 
$$
\l(\f{1}{g_a^2}\r)_{\rm tree}=\f{\pi R}{g_{5a}^2}+
\f{1}{g_{0a}^2}+\f{1}{g_{\pi a}^2}
$$
denote the tree-level couplings and
the one-loop corrections can be written as
\bea
\l(\f{1}{g_a^2}\r)_{\rm loop}&=&
\f{\gamma_a}{24\pi^3}\Lambda\pi R 
+\f{1}{8\pi^2}\l[\tilde{b}_a\ln\Lambda+
\tilde{\Delta}_a(A,B_0,B_\pi,C,k,R)-b_a\ln p\,\r]\,,
\nonumber \\
&\equiv& \f{\gamma_a}{24\pi^3}\Lambda\pi R
+\f{1}{8\pi^2}\l[\,\Delta_a(\ln\Lambda,A,B_0,B_\pi,C,k,R)
+b_a\ln\l(\f{\Lambda}{p}\r)\,\r]\,,
\eea
where the cutoff scale $\Lambda$ is assumed to be large enough compared
to other mass parameters of the theory, so the parts
suppressed by $1/\Lambda$ are ignored.
The linearly divergent one-loop corrections 
are UV-sensitive, i.e. regularization scheme-dependent,
thus can {\it not} be computed within our orbifold field theory
to a level better than just constraining them by 5D gauge symmetry. 
On the other hand, the logarithmically divergent or finite corrections
are UV-insensitive, thus can be computed unambiguously within
orbifold field theory.
Note that the coefficients of $\ln p$, i.e. $b_a$,
correspond to the standard one-loop beta function
coefficients which are determined by the massless spectrum.

One may rewrite the one-loop corrections as
$$
\l(\f{1}{g_a^2}\r)_{\rm loop}=
\Delta_a'+\f{b_a}{8\pi^2}\ln\l(
\f{M_{KK}}{p}\r)\,,
$$
where $M_{KK}$ denotes the mass of the {\it lightest} KK state
which is still bigger than $p$.
Then 
$$
\Delta_a'=
\f{\gamma_a}{24\pi^3}\Lambda\pi R
+\tilde{\Delta}_a+\tilde{b}_a\ln\Lambda-b_a\ln M_{KK}
=
\f{\gamma_a}{24\pi^3}\Lambda\pi R
+\Delta_a+b_a\ln(\Lambda/M_{KK})
$$
could be interpreted as the full threshold corrections
due to the KK modes at scales between
$\Lambda$ and $M_{KK}$.
However $\Delta'_a$ contain the non-calculable
power-law divergences. Also, since $M_{KK}$ is a 
non-trivial function of $R$ and the fundamental mass parameters of
the model, $\Delta_a'$ do {\it not} represent directly
the dependence of $g_a^2(p)$ on $R$ and the
fundamental parameters of the model.
It is thus more convenient
to parameterize $g_a^2(p)$ as
\bea
\label{gaugecoupling}
\f{1}{g_a^2(p)}=\l(\f{1}{g_a^2}\r)_{\rm bare}+
\f{1}{8\pi^2}\l[\Delta_a(\ln\Lambda, A,B_0,B_\pi,C,k,R)+
b_a\ln\l(\f{\Lambda}{p}\r)\r]\,,
\eea
where all uncalculable parts are encoded in
$$
\l(\f{1}{g_a^2}\r)_{\rm bare}=\f{\pi R}{g_{5a}^2}
+\f{1}{g_{0a}^2}+\f{1}{g_{5a}^2}+
\f{\gamma_a}{24\pi^3}\Lambda\pi R
$$ 
and all calculable dependences of $g_a^2(p)$ on
$R$ and the fundamental mass parameters are encoded in
$\Delta_a$.
In the following, we will simply call $\Delta_a$ the
KK threshold corrections.

In orbifold GUT, the {\it unified} higher dimensional gauge symmetry
$G_{GUT}$ assures that both $g_{5a}^2$ and 
the power-law divergent corrections $\gamma_a\Lambda$ are
{\it universal}. On the other hand, since $G_{GUT}$ is
generically broken at the fixed points by boundary condition, 
the bare fixed point gauge
couplings, $g_{0a}^2$ and $g_{\pi a}^2$, are neither universal
nor calculable. However in models with an orbifold radius 
significantly larger than the cutoff length scale,
i.e. $R\Lambda\gg 1$, which would be required for 5D orbifold
field theory to be a useful theoretical framework,
we have $g_{5a}^2\approx \pi R g_a^2 \gg 1/\Lambda$,
implying that the theory is strongly coupled at the cutoff scale
$\Lambda$. A simple naive dimensional analysis suggests
that the most plausible parameter region is given by \cite{luty}
$$
\f{1}{g_{5a}^2}\approx \f{1}{\pi R}={\cal O}\l(\f{\Lambda}{24\pi^3}
\r)
\,,\quad
\f{1}{g_{0a}^2}\approx \f{1}{g_{\pi a}^2}=
{\cal O}\l(\f{1}{8\pi^2}\r)\,.
$$
In this strong coupling limit, the uncalculable fixed point
gauge couplings can be safely ignored, yielding
\beq
\l(\f{1}{g_a^2}\r)_{\rm bare}=\f{1}{g_{GUT}^2}+{\cal O}\l(
\f{1}{8\pi^2}\r)\,,
\eeq
and then the differences between low energy gauge couplings are
dominated by the renormalization group (RG)
running due to zero modes and also the calculable
KK threshold corrections $\Delta_a$.

The computation of KK threshold corrections
involves the summation over all massive
KK modes.
However the involved KK summation can be replaced
by a contour integration with a pole function $P(q)$
which has (simple) poles at $q=m_n$ where $\{m_n\}$ denote
the KK mass eigenvalues \cite{nib}.
The pole function we will use here is given by 
\beq 
P(q) = \f{N'(q)}{2 N(q) }\,,
\eeq
where $N(q)$ has zeroes at $ q=m_n$.
Then the summation over KK modes can be replaced by a counter integral
\beq
\sum_{m_n}\int d^4p \,
f(p, m_n)=\int \frac{dq}{2\pi i}\int d^4p\,  \frac{N'(q)}{2N(q)}
f(p,q)\,,
\eeq
which allows us to compute  
$\Delta_a$ without having  the detailed knowledge of $\{m_n\}$.
The actual computation of $\Delta_a$ using this prescription is
somewhat tedius, but still straightforward.
By computing the one-loop effective action of gauge field
zero modes in this approach, we find \cite{choi2}
\bea
\label{result1}
\Delta_a
&=&-\f{1}{6} T_a(\phi^{(0)}_{++}) \l[\, \ln R_{++}+\ln(\Lm/k)+\pi kR\, \r]
\nonumber \\
&-&\f{1}{6} T_a(\phi_{++}) \l[\,
\ln Q_{++}-\ln(\Lm/k)\, \r]\,
\nonumber \\
&-&\f{1}{6}T_a(\phi_{+-}) \ln Q_{+-} 
-\f{1}{6} T_a(\phi_{-+})\ln Q_{-+} \,
\nonumber \\
&-&\f{1}{6} T_a (\phi_{--}) \l[\, \ln Q_{--}+\ln(\Lm/k)
\,\r]\,
\nonumber \\
&-&\f{2}{3} T_a (\Psi_{++})  \l[\,\ln(\Lambda/k) +\f{1}{2}\pi k R   
+\ln \l\{\f{\displaystyle e^{\l(C_{++}-\f{1}{2}\r)\pi k R} 
-e^{-\l(C_{++}-\f{1}{2}\r)\pi k R}}
{2\l(C_{++}-\f{1}{2}\r)} \r\} \,\r]
\,\nonumber \\
&+&
\f{2}{3} T_a (\Psi_{+-})  C_{+-}\pi k R 
-\f{2}{3} T_a (\Psi_{-+}) C_{-+}\pi k R 
\,\nonumber \\
&-&\f{2}{3} T_a (\Psi_{--})  \l[\,\ln(\Lambda/k) +\f{1}{2}\pi k R   
+\ln \l\{ \f{\displaystyle e^{\l(C_{--}+\f{1}{2}\r)\pi k R}
 -e^{-\l(C_{--}+\f{1}{2}\r)\pi k R}}{2
\l(C_{--}+\f{1}{2}\r)} \r\} \,
\r]
\,\nonumber \\
&+&
\f{1}{12} T_a(A^M_{++}) \l[\, 21 \ln (\Lm \pi R) 
 + 22 \pi k R  \,
\r]\,\nonumber \\
&-&\f{11}{6} T_a(A^M_{+-}) \pi k R\,
+\f{11}{6} T_a(A^M_{-+}) \pi k R\,
\nonumber \\  
&+&\f{1}{12} T_a(A^M_{--}) \l[ 21 \ln (\Lm \pi R) 
 - \pi k R  
+ 21 \ln \l( \f{\displaystyle e^{\pi k R}- e^{-\pi kR}}{2\pi kR} \r)
\r]\,,
\eea
where the subscripts $\pm$ represent 
the $Z_2\times Z_2'$ boundary conditions, 
 $T_a(\Phi)=\mbox{Tr}(T_a^2(\Phi))$ is the Dynkin index
of the gauge group representation $\Phi$,
and $C_{zz'}k$ is the kink mass of $\Psi_{zz'}$. 
Here $\phi^{(0)}_{++}$ denotes a 5D complex scalar field having 
a zero mode, i.e. a scalar field whose bulk and brane
mass parameters (see Eq.(\ref{mass})) satisfy 
\bea
&&B_0(\phi^{(0)}_{++})=B_\pi
(\phi^{(0)}_{++})\equiv B_{++}\,,
\nonumber \\
&&\sqrt{4+A^2(\phi^{(0)}_{++})}
=|\,2-B_{++}\,|\,,
\eea
while
$\phi_{zz'}$ ($z,z'=\pm 1$) stand for complex scalar fields
without zero mode.
The functions that appear in $\Delta_a$ for
5D scalar fields are given by
\bea
Q_{++} &=&
\f{1}{2\a_{++}}\l[\, 
(\a_{++} +B_{0++}-2)(\a_{++}-B_{\pi ++}+2)e^{\a_{++} \pi kR}
\r.\nonumber \\
&&\l.\quad\quad\quad
-(\a_{++}+B_{\pi ++}-2)(\a_{++}-B_{0++}+2)e^{-\a_{++} \pi kR} 
\,\r],
\nonumber \\
Q_{+-} &=&
\f{1}{2\a_{+-}}\l[\,
(\a_{+-}+B_{0+-}-2)e^{\a_{+-} \pi kR}+
 (\a_{+-}-B_{0+-}+2)e^{-\a_{+-} \pi kR}
\,\r],
\nonumber \\
Q_{-+} &=& 
\f{1}{2\a_{-+}}\l[\, (\a_{-+}-B_{\pi -+}+2)e^{\a_{-+}\pi kR}+
(\a_{-+}+B_{\pi -+}-2)e^{-\a_{-+}\pi kR}\,\r],
\nonumber \\
Q_{--} &=& 
\f{1}{2 \a_{--}} \l[\, e^{\a_{--} \pi k R} - e^{-\a_{--} \pi k R}\, \r], 
\nonumber \\
R_{++} &=&
\f{1}{2(1-B_{++})}\l[\,e^{(1-B_{++}) \pi kR}-e^{-(1-B_{++})
\pi kR}\,\r]\,,
\nonumber 
\eea
where 
$$\a_{zz'}=
\sqrt{4+A^2_{zz'}}$$ 
and the parameters $A_{zz'}$,$B_{0 zz'}$,$B_{\pi zz'}$
and $C_{zz'}$ ($z,z'=\pm 1$) are defined through the scalar and fermion
masses (see Eq.(\ref{mass})):
$$
m^2_{zz'}=A_{zz'}+\f{2k}{R}\l[B_{0zz'}\delta(y)-B_{\pi zz'}\delta(y-\pi)\r]
\,,
\quad
M_{zz'}=C_{zz'}k.
$$
The one-loop beta function coefficients $b_a$ in 
(\ref{gaugecoupling}) are given by
\beq
b_a=-\f{11}{3}T_a(A^M_{++})+\f{1}{6}T_a(A^M_{--})
+\f{1}{3}T_a({\phi}^{(0)}_{++})+\f{2}{3}T_a(\Psi_{++})
+\f{2}{3}T_a(\Psi_{--}),
\eeq
which can be easily understood by noting that
$A^M_{++}$ gives a massless 4D vector, $A^M_{--}$ gives
a massless real 4D scalar, and $\Psi_{zz'}$ with $z=z'=\pm 1$ gives
a massless 4D chiral fermion.

In the above, we considered only the 
KK threshold corrections which are {\it parametrically enhanced
by the large logarithms of scale ratios}, while ignoring
the scheme-dependent constant parts of order unity.
There appear a variety of logs in $\Delta_a$, from
 $\ln(e^{\omega \pi kR})$
($\omega=1, \a_{zz'},C_{zz'}$) to $\ln(\Lambda/k)$
and $\ln (\omega'\pi kR)$ ($\omega'=1,\a_{zz'},B_{0zz'},
B_{\pi zz'}, C_{zz'})$.
It is obvious that $\Delta_a={\cal O}(\pi MR)$ in general, 
where $e^{-\pi MR}$ corresponds to
either the warp factor or the small Yukawa couplings
of quasi-localized fermions. Thus, 
in orbifold field theories in which the 4D graviton
and/or matter fermions are quasi-localized to generate
hierarchically different scales and/or Yukawa couplings,
we have
\beq
\Delta_a ={\cal O}(\ln e^{-\pi kR})\quad
\mbox{and/or}
\quad {\cal O}(\ln y)\,.
\eeq

The expression of $\Delta_a$ in (\ref{result1}) is based on the assumption
that there exists a large mass gap between the {\it lightest}
KK mass ($M_{KK}$) and the zero mode masses.
Note that all KK states are treated as superheavy, while all 
zero modes are considered to be massless. 
The low energy gauge couplings at 
$p$ below $M_{KK}$ (but above the zero mode masses)
are determined as  (\ref{gaugecoupling}) at one-loop approximation.
In many 5D orbifold field theories, we have
\beq
\label{scalehierarchy}
\f{M_{KK}}{M_W}\,\gg \,\f{\Lambda}{M_{KK}}
\eeq
by many orders of magnitude, where $M_W$ is the weak scale.
Then the dominant part of higher order corrections
(beyond one-loop) to low energy couplings at $M_W$
come from the energy scales below $M_{KK}$, which
can be systematically computed within 4D effective theory.
To include those higher order corrections,
one can start with the matching condition
at $M_{KK}$:
\beq
\f{1}{g_a^2(M_{KK})}=\l(\f{1}{g_a^2}\r)_{\rm bare}
+\f{1}{8\pi^2}\l[
\Delta_a+b_a\ln\l(\f{\Lambda}{M_{KK}}\r)\r]\,,
\eeq
where $\Delta_a$ are given by (\ref{result1}),
and then subsequently perform 
two-loop RG analysis
over the scales between $M_{KK}$ and $M_W$.

In some case, there can be another large mass gap between
the lightest KK mass $M_{KK}$ and the {\it next} lightest
KK mass  ${M}^\pr_{KK}$.
In fact, such additional mass gap is quite common in orbifold
field theories in which the scale and/or Yukawa hierarchies
are generated by the quasi-localization of 4D graviton and/or
fermions.
In such case with
\beq
\label{scalehierarchy2}
\f{M_{KK}}{M_W}\gg
\f{{M}^\pr_{KK}}{M_{KK}}\,\gg\, \f{\Lambda}{{M}^\pr_{KK}}\,,
\eeq
the next important higher order corrections would come
from energy scales between $M_{KK}$ and $M^\pr_{KK}$. 
Those higher order corrections can be included
by performing the two-loop RG analysis starting from $M^\pr_{KK}$.
The corresponding matching condition
at $M_{KK}^\pr$ is given by
\beq
\f{1}{g^2_a(M^\pr_{KK})}=\l(\f{1}{g_a^2}\r)_{\rm bare}
+\f{1}{8\pi^2}\l[\Delta^\pr_a +b_a^\pr \ln\l(\f{\Lambda}{M^\pr_{KK}}
\r)\r]\,,
\eeq
where
\bea
b^\pr_a &=&b_a+\delta b_a\,,
\nonumber \\
\Delta^\pr_a &=& \Delta_a-\delta b_a \ln 
\l(\f{\Lambda}{M_{KK}}\r)
\eea
for $\Delta_a$ given by (\ref{result1}).
Here $b_a$ denote the one-loop beta function
coefficients due to zero modes, while
$\delta b_a$ denote the 
coefficient due to the lightest KK states.

\section{4d supergravity calculation}

In supersymmetric 5D theories, one-loop low energy gauge couplings   
can be computed using the gauged $U(1)_R$ symmetry and chiral 
anomaly of 5D SUGRA on orbifold and also
the known properties of  gauge couplings in 4D effective SUGRA
\cite{choi1}.
In this section, we discuss this alternative way to
compute the KK threshold corrections in supersymmetric theories, 
and show that the SUGRA results agree with the results of the previous
section.
To proceed, let us briefly discuss supersymmetric 5D theory
on $\mbox{AdS}_5$.
The theory contains two types of 5D supermultiplets other than
the SUGRA multiplet.
One is the hypermultiplet
${\cal H}$ containing two 5D complex scalar fields $h^i$ ($i=1,2$) and
a Dirac fermion $\Psi$, and the other is the vector
multiplet ${\cal V}$ containing a 5D vector $A_M$, real scalar $\Sigma$ and
a symplectic Majorana fermion $\lambda^i$.
In supersymmetric model, all 5D scalar fields and their superpartner
fermions have 
$$B_0=B_\pi=B\,,\quad
\sqrt{4+A}=|2-B|\,,\quad
C=\pm (3-2B)/2\,.
$$ 
See Eqs. (\ref{mass})
for the definitions of $A$, $B_{0,\pi}$ and $C$.
Also the $U(1)_R$ symmetry is gauged with the graviphoton $\Omega_M$
in the following way \cite{choi1}:
\bea
&& D_Mh^i =\partial_M h^i-i\left(\frac{3}{2}(\sigma_3)^i_j-C\delta^i_j\right)
k\epsilon(y)\Omega_M h^j +...\nonumber \\
&& D_M\Psi=\partial_M \Psi +iCk\epsilon(y)\Omega_M\Psi+...
\nonumber \\
&& D_M\lambda^{i}=\partial_M\lambda^{i}-i\frac{3}{2}(\sigma_3)^i_j k
\epsilon(y)\Omega_M\lambda^{j}+...\,,
\eea
where $\Psi$ has a kink mass $Ck\epsilon(y)$ and
the ellipses stand for the couplings with other gauge fields.
Taking into account the $Z_2\times Z^{\prime}_2$ parity, the
supermultiplet structure is given by
\bea
&&{\cal H}_{zz^{\prime}}(C)=\l(\,h^1_{zz^{\prime}}(B=\f{3}{2}-C)\,,\,
 h^2_{\tilde{z}\tilde{z}^{\prime}}(B=\f{3}{2}+C)\,,\, 
\Psi_{zz^{\prime}}(C)\,\r)\,,\nonumber \\
&&{\cal V}_{zz^{\prime}}=\l(\, 
A^M_{zz'}=(A^\mu_{zz^{\prime}}, 
A^5_{\t{z}\t{z}^{\prime}}(B=2))\,,\, 
\lambda^i=\lambda^{\rm Dirac}_{zz^{\prime}}(C=\f{1}{2})\,,\,
\Sigma_{\t{z}\t{z}^{\prime}}(B=2)\,\r)\,,
\eea
where $z,z^{\prime}=\pm 1$, 
$\t{z}=-z$, $\t{z}^{\prime}=-z^{\prime}$,
$B$ is the fixed point mass parameter
and $C$ is the kink mass parameter.

Let us assume that our 5D theory is compactified
in a manner preserving $D=4$ $N=1$ supersymmetry.
Then the low energy physics can be described by
a 4D effective SUGRA action which can be written as
\beq \label{4daction} S_{4D}=\int d^4x \, \left[\,\int
d^4\theta  \, \left\{-3\exp \left(-\frac{K}{3}\right)\right\}
+\left( \int d^2\theta \, \frac{1}{4}
f_{a}\,W^{a\alpha}W^a_{\alpha}+h.c.\,\right)\,\right]\,, \eeq
where $W^a_{\alpha}$ is the chiral spinor superfield for the
4D gauge multiplet and the 4D SUGRA multiplet is replaced
by their vacuum expectation values. 
The K\"ahler potential $K$ can be expanded in powers of
generic gauge-charged chiral superfield $Q$: \beq K\,=\,
K_0(T,T^*)+Z_{Q}(T,T^*)Q^*e^{-V}Q+...\,, \eeq where $T$
denotes the radion superfield whose scalar component
is given by
$$
T=R+i\Omega_5\,,
$$
where $\Omega_5$ is the fifth-component of the 5D
graviphoton,
and the gauge kinetic function $f_a$ is  a 
{\it holomorphic} function of $T$.
Then  the one-loop gauge couplings in effective 4D SUGRA
can be determined by
$f_a$ containing the one-loop
threshold corrections from massive KK modes and also
the tree-level K\"ahler potential $K$ \cite{kaplunovsky}:
\bea
\label{4dsugracoupling} \frac{1}{g^2_a(p)}\,&=&\,
\mbox{Re}(f_a)+\frac{b_a}{16\pi^2}
\ln\left(\frac{M_{Pl}^2}{e^{-K_0/3}p^2}\right) \nonumber \\
&&-\sum_{Q}\frac{T_a(Q)}{8\pi^2}\ln\left(e^{-K_0/3}Z_{Q}
\right) +\frac{T_a({\rm Adj})}{8\pi^2}\ln\left({\rm Re}(f_a)\right), \eea
where $b_a=\sum T_a(Q)-3T_a({\rm Adj})$ are the one-loop beta
function coefficients and $M_{Pl}$ is the Planck scale of
$g_{\mu\nu}$ which defines the momentum scale
$p^2=-g^{\mu\nu}\partial_\mu\partial_\nu$.

Let us consider the 4D effective SUGRA of a 5D theory
which contains generic 5D hypermultiplets and vector multiplets,
${\cal H}_{zz'}$ and
${\cal V}_{zz'}$, with arbitrary
boundary conditions.
The vector multiplet
${\cal V}_{++}$ gives a massless 4D gauge multiplet
containing $A^\mu_{++}$ whose 
low energy couplings are of interest for us, while
${\cal V}_{--}$ gives a massless
4D chiral multiplet containing $\Sigma_{++}+iA^5_{++}$. 
${\cal H}_{++}$ and ${\cal H}_{--}$ also give massless 4D chiral multiplets
containing $h^1_{++}$ and $h^2_{++}$, respectively,
whose tree level K\"ahler metrics
are required to compute the one-loop gauge couplings (\ref{4dsugracoupling}).
Other multiplets, i.e. ${\cal V}_{+-},{\cal V}_{-+},{\cal H}_{+-}$ and
${\cal H}_{-+}$ do not give any massless 4D mode.

Let $Z_Q$ 
($Q={\cal H}_{++},{\cal H}_{--},{\cal V}_{--}$) denote the 
K\"ahler metric of the 4D massless chiral superfields coming from
the 5D multiplets ${\cal H}_{++}, {\cal H}_{--}$ and ${\cal V}_{--}$.
It is then straightforward
to compute the {\it tree level} $Z_Q$
and also $f_a$ containing {\it the 1-loop threshold corrections}
from massive KK modes: 
\bea \label{kahler} 
&&
M_{Pl}^2=e^{-K_0/3}\Lambda^2=\f{M_5^3}{k}(1-e^{-k\pi(T+T^*)})
\,,\nonumber\\
&& e^{-K_0/3}Z_{{\cal H}_{++}}\,=\,
\frac{\Lambda}{(\frac{1}{2}-C_{++})k} (e^{(\frac{1}{2}-C_{++})\pi
k(T+T^*)}-1)\,, \nonumber \\
&& e^{-K_0/3}Z_{{\cal H}_{--}}\,=\,\f{\Lambda}{(\f{1}{2}+C_{--})k}
(e^{(\f{1}{2}+C_{--})\pi k(T+T^*)}-1)\,,
\nonumber \\
&& e^{-K_0/3}Z_{{\cal V}_{--}}\,=\,\f{k }{\Lambda} 
\f{1}{e^{\pi k (T+T^*)} -1}\,,
\nonumber \\ 
&& f_{a}\,=\,\frac{\pi T}{{g}^2_{5a}} +\frac{z^{\prime}}{8\pi^2}\left(
\frac{3}{2}\sum_{{\cal V}_{zz^{\prime}}}T_a({\cal V}_{zz^{\prime}})-
\sum_{H_{zz^{\prime}}}C_{zz^{\prime}}
T_a({\cal H}_{zz^{\prime}})\right)
k\pi T\,, \eea 
where $\Lambda$ and $M_5$ are the 5D cutoff scale and the 5D Planck scale,
respectively, and $C_{zz^\prime}$ is the kink mass of ${\cal H}_{zz^\prime}$.
The scale ratio $\Lambda/M_5$ is not sensitive to
the values of relevant physical variables like
$C_{zz'}$, $k$ and $R$, so can be set to be a constant of order unity,
$\Lambda/M_5\approx 1$.
The KK threshold correction to $f_a$
can be entirely determined by the chiral anomaly w.r.t the
following $\Omega_5$-dependent phase transformation:
\bea \label{phase} &&
\lambda^{ai} \rightarrow 
\l(e^{3iky\Omega_5\sigma_3/2}\r)^i_j \lambda^{aj}\,, \quad
\Psi \rightarrow e^{-iCky\Omega_5}\Psi\,. 
\eea
Using the above results, we find \cite{choi1}
\bea
\label{4dbulkcoupling}
\l(\Delta_a\r)_{\rm SUSY}=&&-T_a({\cal H}_{++})\left[\,\ln
\left(\frac{\Lambda}{k}\right)+C_{++} \pi kR 
+\ln \left(\f{
e^{(1-2C_{++})\pi kR}-1}{1-2C_{++}}\right)\,\right] 
\nonumber \\
&&+C_{+-}T_a({\cal H}_{+-})\pi kR
-C_{-+}T_a({\cal H}_{-+})\pi kR
\nonumber \\
&&-T_a({\cal H}_{--})\l[ \,\ln \l(\f{\Lambda}{k}\r)-C_{--}\pi kR
+\ln\l(\f{e^{(1+2C_{--})\pi kR}-1}{1+2C_{--}}
\r)\,\r]
\nonumber \\
&& +T_a({\cal V}_{++})\left[\,\ln (\Lambda \pi R)+\f{3}{2}\pi k R 
\,\right]
 \nonumber \\
&&-\f{3}{2}T_a({\cal V}_{+-})\pi kR
+\f{3}{2}T_a({\cal V}_{-+})\pi kR
\nonumber \\
&&+T_a ({\cal V}_{--}) \l[ \, \ln \f{\Lambda}{k} +\f{1}{2} k \pi R +
\ln \l(\f{1- e^{-2\pi kR}}{2}\r)\, \r]\,
\eea
and also the 4D beta function coefficients
$$
\l(b_a\r)_{\rm SUSY}=
-3T_a({\cal V}_{++})+
T_a({\cal V}_{--})+T_a({\cal H}_{++})+
T_a({\cal H}_{--}).
$$
The above result obtained by 4D SUGRA analysis perfectly agrees with
the results of the previous section
in supersymmetric limit.
This provides a nontrivial
check for the results of the previous section and
assures that our results are regularization scheme-independent.

\section{results for flat extra dimension}

In fact, most of the 5D theories using the quasi-localization
of fermions to generate hierarchical Yukawa couplings have been
constructed on a {\it flat} orbifold, not on a slice of $\mbox{AdS}_5$.
We thus summarize seperately the KK threshold corrections in theories
on flat orbifold, which
can be obtained from the results for
$\mbox{AdS}_5$ by taking the limit $k\rightarrow 0$,
while keeping the scalar and fermion mass parameters,
$Ak^2, B_{0,\pi}k$ and $Ck$, to be  non-vanishing \cite{choi3}:
\bea
\label{result2}
\Delta_a =&&
-\f{1}{6}T_a({\phi^{(0)}}_{++})\ln \l(\f{\Lambda(e^{m_{++}\pi R}
-e^{-m_{++} \pi R})}{2m_{++}}\r)
\nonumber \\
&&-\f{1}{6}T_a(\phi_{++})\ln\l(\f{(m_{++}+\mu_{++})(m_{++}-\mu^\pr_{++})e^{m_{++}\pi R}
-(m_{++}-\mu_{++})(m_{++}+\mu^\pr_{++})e^{-m_{++}\pi R}}{2m_{++}\Lambda}\r)
\nonumber \\
&&-\f{1}{6}T_a(\phi_{+-})\ln\l(
\f{(m_{+-}+\mu_{+-})e^{m_{+-}\pi R}+(m_{+-}-\mu_{+-})e^{-m_{+-}\pi R}}{2m_{+-}}\r)
\nonumber \\
&&-\f{1}{6}T_a(\phi_{-+})\ln\l(
\f{(m_{-+}-\mu^\pr_{-+})e^{m_{-+}\pi R}+(m_{-+}+\mu^\pr_{-+})e^{-m_{-+}\pi R}}{2m_{-+}}\r)
\nonumber \\
&&-\f{1}{6}T_a(\phi_{--})\ln\l(
\f{\Lambda (e^{m_{--}\pi R}-e^{-m_{--}\pi R})}{2m_{--}}\r)
\nonumber \\
&&-\f{2}{3}T_a(\Psi_{++})\ln\l(\f{\Lambda(e^{{M}_{++}\pi R}
-e^{-{M}_{++}\pi R})}{2{M}_{++}}\r)
\nonumber \\
&&-\f{2}{3}T_a(\Psi_{+-})\ln \l(e^{-{M}_{+-}\pi R}\r)
\nonumber \\
&&-
\f{2}{3}T_a(\Psi_{-+})\ln\l(e^{{M}_{-+}\pi R}\r)
\nonumber \\
&&-\f{2}{3}T_a(\Psi_{--})\ln\l(
\f{\Lambda (e^{{M}_{--}\pi R}-e^{-{M}_{--}\pi R})}
{2{M}_{--}}\r)\,
\nonumber \\
&&+\frac{21}{12}\l[T_a(A^M_{++})+T_a(A^M_{--})\r]\ln(\Lambda\pi R)
\eea
where $m_{zz'}, \mu_{zz'}$ and  $\mu^\pr_{zz'}$
denote the bulk and brane masses of $\phi_{zz'}$,
$$
\hat{m}^2=m^2+\f{2}{R}\l[\,
\mu\delta(y)-\mu'\delta(y-\pi)\,\r]\,
$$
and ${M}_{zz'}$ is the kink mass of $\Psi_{zz'}$.
Again, ${\phi^{(0)}}_{++}$ is a 5D scalar field
having a zero mode, i.e. a scalar field with
$\mu=\mu^\pr=m$, and $\phi^{++}$ stands for scalar fields
{\it without} zero mode.
Similarly, in supersymmetric case, we have
\bea
\label{result3}
\l(\Delta_a\r)_{\rm SUSY}=&&
-T_a({\cal H}_{++})\ln \l(\f{\Lambda(e^{{M}_{++}\pi R}
-e^{-{M}_{++}\pi R})}{2{M}_{++}}\r)
\nonumber \\
&&-T_a({\cal H}_{+-})\ln \l(e^{-{M}_{+-}\pi R}\r)
\nonumber \\
&&-T_a({\cal H}_{-+})\ln \l(e^{{M}_{-+}\pi R}\r)
\nonumber \\
&&-T_a({\cal H}_{--})\ln \l(\f{\Lambda(e^{{M}_{--}\pi R}
-e^{-{M}_{--}\pi R})}{2{M}_{--}}\r)
\nonumber \\
&&+
T_a({\cal V}_{++})\ln (\Lambda\pi R)
+ T_a({\cal V}_{--})\ln (\Lambda \pi R)\,,
\eea
where $M_{zz'}$ is the kink mass of the hypermultiplet
${\cal H}_{zz'}$.



\bigskip
{\bf Acknowledgements}

\medskip

This work is supported by KRF PBRG 2002-070-C00022 and
KOSEF through Center for High Energy Physics of Kyungpook
National University.


\begin{thebibliography}{99}



\bibitem{add}
N. Arkani-Hamed, S. Dimopoulos and G. R. Dvali, Phys.\ Lett. {\bf B429},
263 (1998).

\bibitem{rs}
L.~Randall and R.~Sundrum,
Phys.\ Rev.\ Lett.\  {\bf 83}, 3370 (1999);
Phys. Rev. Lett. {\bf 83}, 4690 (1999).

\bibitem{yukawa}
N.~Arkani-Hamed and M.~Schmaltz,
Phys.\ Rev.\ D {\bf 61}, 033005 (2000);
E.~A.~Mirabelli and M.~Schmaltz,
Phys.\ Rev.\ D {\bf 61}, 113011 (2000);
D.~E.~Kaplan and T.~M.~Tait,
JHEP {\bf 0111}, 051 (2001);
M.~Kakizaki and M.~Yamaguchi,
arXiv:hep-ph/0110266;
N.~Haba and N.~Maru,
Phys.\ Rev.\ D {\bf 66}, 055005 (2002);
Y.~Grossman and G.~Perez,
Phys.\ Rev.\ D {\bf 67}, 015011 (2003);
K.~Choi, D.~Y.~Kim, I.~W.~Kim and T.~Kobayashi,
hep-ph/0305024.


\bibitem{kawamura}
Y.~Kawamura,
Prog.\ Theor.\ Phys.\  {\bf 105}, 691 (2001);
Prog. Theor. Phys. {\bf 105}, 999 (2001);
A.~Hebecker and J.~March-Russell,
Nucl.\ Phys.\ B {\bf 613}, 3 (2001);
Nucl.\ Phys.\ B {\bf 625}, 128 (2002);
G.~Altarelli and F.~Feruglio,
Phys.\ Lett.\ B {\bf 511}, 257 (2001);
L.~J.~Hall and Y.~Nomura,
Phys.\ Rev.\ D {\bf 64}, 055003 (2001);
Phys.\ Rev.\ D {\bf 66}, 075004 (2002);
H.~D.~Kim, J.~E.~Kim and H.~M.~Lee,
JHEP {\bf 0206}, 048 (2002);
H.~D.~Kim and S.~Raby,
JHEP {\bf 0301}, 056 (2003);
JHEP {\bf 0307}, 014 (2003).




\bibitem{hebecker}
A.~Hebecker and J.~March-Russell,
Phys.\ Lett.\ B {\bf 541}, 338 (2002);
A. Hebecker, J. March-Russell and T. Yanagida,
Phys. Lett. {\bf B552}, 229 (2003);
R.~Kitano and T.~j.~Li,
Phys.\ Rev.\ D {\bf 67}, 116004 (2003).


\bibitem{choi}
K.~Choi,
Phys.\ Rev.\ D {\bf 37}, 1564 (1988);
V.~S.~Kaplunovsky,
Nucl.\ Phys.\ B {\bf 307}, 145 (1988)
[Erratum-ibid.\ B {\bf 382}, 436 (1992)].






\bibitem{ddg}
K. R. Dienes, E. Dudas and T. Gherghetta,
Phys. Lett. {\bf B436}, 55 (1998);
Nucl. Phys. {\bf B537} 47 (19990.


\bibitem{contino}
R.~Contino, L.~Pilo, R.~Rattazzi and E.~Trincherini,
Nucl.\ Phys.\ B {\bf 622}, 227 (2002).

\bibitem{hall}
L. J. Hall and Y. Nomura, 
Phys. Rev. {\bf D65}, 125012 (2002).


\bibitem{powerlaw}
A. Hebecker and A. Westphal, Annals Phys. {\bf 305}, 119
(2003);
J.F. Oliver, J. Papavassiliou and
A. Santamaria,
Phys. Rev. {\bf D67}, 125004 (2003);






\bibitem{choi1}
K.~Choi, H.~D.~Kim and I.~W.~Kim,
JHEP {\bf 0211}, 033 (2002);
JHEP {\bf 0303}, 034 (2003);


\bibitem{pomarol}
A.~Pomarol,
Phys.\ Rev.\ Lett.\  {\bf 85}, 4004 (2000);
L.~Randall and M.~D.~Schwartz,
JHEP {\bf 0111}, 003 (2001);
Phys.\ Rev.\ Lett.\  {\bf 88}, 081801 (2002).




\bibitem{goldberger}
W. D. Goldberger and Ira Z. Rothstein,
Phys. Rev. Lett. {\bf 89}, 131601 (2002); hep-ph/0303158;
hep-th/0208060;  K. Agashe, A. Delgado and R. Sundrum,
 Nucl. Phys. {\bf B643}, 172 (2002); Annals Phys. {\bf 304}, 145
(2003); R. Contino, P. Creminelli and E. Trincherini,
JHEP {\bf 0210}, 029 (2002).




\bibitem{choi2}
K.~Choi and I.~W.~Kim,
Phys.\ Rev.\ D {\bf 67}, 045005 (2003).






\bibitem{choi3}
K. Choi, I. W. Kim and W. Y. Song, hep-ph/0307365.





\bibitem{luty}
Z.~Chacko, M.~A.~Luty and E.~Ponton,
JHEP {\bf 0007}, 036 (2000);
Y.~Nomura,
Phys.\ Rev.\ D {\bf 65}, 085036 (2002).

\bibitem{nib}
S.~Groot Nibbelink,
Nucl.\ Phys.\ B {\bf 619}, 373 (2001);
R.~Contino and A.~Gambassi,
J.\ Math.\ Phys.\  {\bf 44}, 570 (2003).



\bibitem{kaplunovsky}
V.~Kaplunovsky and J.~Louis,
Nucl.\ Phys.\ B {\bf 422}, 57 (1994).




\end{thebibliography}
\end{document}